\documentclass[twocolumn,showpacs,aps]{revtex4}

\usepackage{graphicx}
\usepackage{pstricks}
\usepackage{dcolumn}
\usepackage{bm}
\usepackage{epsf}
\usepackage{epsfig}
\usepackage{amsmath,amsfonts,amsthm, amssymb,latexsym}
\usepackage{enumerate}

\newcommand{\beq}{\begin{equation}}
\newcommand{\eeq}{\end{equation}}
\newcommand{\bcn}{\begin{center}}
\newcommand{\ecn}{\end{center}}

\newcommand{\lsim}{\lower0.5ex\hbox{$\; \buildrel < \over \sim \;$}}

\begin{document}

\title{Quark core impact on hybrid star cooling}

\author{Rodrigo Negreiros}
 \email{negreiros@fias.uni-frankfurt.de}
\affiliation{FIAS, Goethe University, Ruth Moufang Str. 1
        60438 Frankfurt am Main, Germany}
        
 \author{V.A. Dexheimer}
 \email{vantoche@gettysburg.edu}
\affiliation{Gettysburg College, Gettysburg, USA}

\author{S. Schramm}
 \email{schramm@th.physik.uni-frankfurt.de}
\affiliation{CSC, FIAS, ITP, Johann Wolfgang Goethe University, Frankfurt am
Main, Germany}

\date{\today}

\begin{abstract}
In this paper we investigate the thermal evolution of hybrid stars, objects
composed of a quark matter core, enveloped by  ordinary hadronic matter.
Our purpose is to investigate how important are the  microscopic properties of the 
quark core to the thermal evolution of the star. In order to do that we
use a simple MIT bag model for the quark core, and a relativistic mean field
model for the hadronic envelope. By choosing different values for the microscopic parameters (bag constant, strange quark mass, strong coupling constant)  we obtain hybrid stars with different quark core properties. We also consider the possibility of color superconductivity in the quark core.  With this simple approach, we have found a set of microscopic parameters that lead to a good agreement with observed cooling neutron stars. Our results can be used to obtain clues regarding the properties of the quark core in hybrid stars, and can be used to refine more sophisticated models for the equation of state of quark matter.
\end{abstract}

\maketitle

\section{Introduction} \label{sec:Int}
Scientists have been trying to determine the equation of state and composition
of compact stars for a long time (\cite{Weber,Glendenning2000} and references
therein). Many of these studies determine constraints for the equation of state
of stellar matter so that the predicted stellar properties agree with the
observed masses and radii of pulsars. One of the biggest practical shortcomings of such
techniques is the lack of accurate and reliable observations for the radii of
compact stars, which makes constraining the equation of state very
challenging. A recent paper \cite{Ozel2010} has put forth unprecedented
constraints on the mass-radius diagrams of pulsars, which would certainly make the range of possible
equations of state narrower. The analysis of this data however has been questioned
\cite{Steiner2010}, and therefore is not universally accepted. Another method
that might be used to complement the results provided by the analysis of the
structure of the compact star is the examination of the thermal evolution of the
object. The cooling of a compact star is intrinsically connected to its
microscopic composition, and therefore different microscopic models might lead
to different thermal evolutions. The results of the theoretical investigations
might then be compared with the wealth of observed data of the thermal
properties of compact stars \cite{Page2004,Page2009}. In this paper we will
follow the thermal evolution approach, and with this probe the composition of
compact stars.

In the following we will consider the thermal evolution of hybrid stars (HS), which
are objects composed of a strange quark matter core, enveloped by ordinary hadronic
matter \cite{Weber,Glendenning2000}. 
In this work, neutron stars (NS) denote objects made up
solely of hadronic matter.

The cooling of neutron stars is dominated by neutrino emission for the first
1000 years (possibly more for the slow cooling scenario) \cite{Page2006},
being replaced by surface photon emission after that. Amongst the neutrino
emission mechanisms, the direct Urca (DU hereafter) process sets itself apart by
being one of the most efficient ones, with emissivity of the order of $10^{26}$
erg cm$^{-3}$ s$^{-1}$ \cite{Lattimer1991}.
Due to momentum conservation, the DU
process can only take place if the proton fraction is above a certain value. The actual
proton-fraction, above which the DU process is allowed to take place, depends on the underlying
microscopic model, and it usually lies between 11\%$-$15\%. If a neutron star allows the DU
process to take place, the star will exhibit a fast cooling (sometimes also
referred to as enhanced
cooling).  If  neutron superfluidity \cite{Schaab94,Levenfish94}
(superconductivity in the case of protons) occurs, pairing will suppress
the emission of neutrinos from hadronic processes and smooth the so-called
enhanced
cooling.
The situation for hybrid stars is significantly different, since
the presence of deconfined quark matter plays an important role for the
cooling of the object. The cooling of hybrid stars has been addressed by many
authors \cite{Blaschke2000,Grigorian2005,Blaschke2006}.
In this work, differently than in references \cite{Blaschke2000,Grigorian2005,Blaschke2006}, where
the cooling of hybrid stars with the same equation of state was studied,  we will examine
the cooling of hybrid stars with different EoS's for the quark phase.  As a
result, we will obtain quark cores with different properties (mass, radius,
composition). This should
provide us
with a measure of how important  such properties are for the 
thermal evolution of hybrid stars, which might aid us in
constraining the EoS of compact stars. In order to perform such study we will
use a MIT bag model equation of state for the quark phase, and a 
relativistic-mean-field model EoS for the hadronic phase.  The different EoS's for the
quark
phase are obtained by varying the microscopic parameters of quark matter, namely the
bag constant $B$, the strong interaction constant $\alpha_s$, and the strange quark mass
$m_s$.  The macroscopic
structure of the star is obtained by solving the Tolman-Oppenheimer-Volkoff
equations \cite{Tolman1939,Oppenheimer1939}. We note that there are more
sophisticated models for hybrid stars, see for
example \cite{Grigorian2004,Negreiros2010,Dexheimer2010}.  However, the
relatively simple MIT bag model is appropriate for our study. Due to
its phenomenological nature, it
permits one to change its microscopic properties with ease, allowing us to
obtain quark cores with
different properties. This in turn allows us to investigate how such quark cores effectively
influence the thermal evolution of hybrid stars. The aim of this procedure is to obtain constraints
on the properties of cold quark matter, which then can be used in more
sophisticated microscopic
calculations.

This paper has the following structure: in Sec.~\ref{Model} we describe the
microscopic model used, and present the results for the macroscopic structure
of the stars; in Sec.~\ref{Results} our results for the thermal
evolution of the stars discussed in Sec.~\ref{Model} are presented; in
Sec.~\ref{SC} the
role of superconductivity in the quark core is discussed, and the results for
the cooling of superconducting hybrid stars are presented; our final
remarks and conclusions can be found in Sec.~\ref{conc}

\section{Model} \label{Model}

We use a standard relativistic mean field model \cite{Weber,Glendenning2000} to describe
the hadronic phase, in which the interactions between the baryons are mediated by meson fields
$(\sigma, \omega, \vec{\rho})$. The parametrization used is G300, which can
successfully reproduce the properties of saturated nuclear matter
\cite{Glendenning1989}.  We have also allowed pairing in the hadronic phase. The pairing
patterns included are the neutron singlet ($^1S_0$) and triplet ($^3P_2$) states, as described in \cite{Levenfish94,Schaab94}. It is also possible that protons can
form pairs, becoming a singlet superconductor;  however, the pairing of
protons in the core is still not very well understood (see \cite{Page2011b} and
references therein). For this reason have decided to take a conservative
approach and only consider neutron
superfluidity.

For the quark matter description we use the MIT Bag Model,
with first order corrections in the strong interaction
coupling constant
($\alpha_s$) \cite{Chodos1974,Chodos1974a,Farhi1984,Weber2005}. In the framework
of the MIT Bag Model, the quarks present at the relevant densities are the up, down and strange
quarks. It
is important to note that, since we are considering massive strange quarks, electrons are required
for charge neutrality to be achieved. At lower densities, the electron population
is more significant, due to suppression of the massive strange quarks. This is
an important phenomenon, that may lead to the formation of ultra-strong electric
fields near the surface of bare strange stars
\cite{Alcock1986,Usov2004,Negreiros2009, Negreiros2010b}. 

In this work we also take into account the possibility of color
superconductivity. The pattern that will be considered is the
Color-Flavor-Locked phase \cite{Alford2001},
where all quarks are paired. The CFL phase is the most
likely condensation pattern at densities $> 2\rho_0$  \cite{Alford2008}. 
For intermediate densities ($\sim 2\rho_0$) model calculations seem to indicate
that quark matter is in a 2SC phase \cite{Alford2008}. Another possibility is
that quark matter forms a crystalline superconductor, where the momenta of the
quark pairs do not add to zero \cite{Alford2001a,Bowers2002}.
Since we are considering quark matter only at the inner core of compact stars,
we will consider only the CFL phase. It should be noted that one expects
corrections to the quark matter EoS if pairing is considered.
As shown in \cite{Lugones2002}, such corrections are $\sim \Delta^2 \mu^2$. The
authors in ref.~\cite{Alford2003} have shown, however, that the effects of such
corrections to the structure of the star are only noticeable for $\Delta
\gtrsim 50$ MeV. Therefore, for the values of $\Delta$ considered in this study
($0.1-10$ MeV) they can be safely disregarded.  Note that as the small values for the gap
($\Delta$) do not have any appreciable effect on the equation of state, our analysis is still
valid for any quark pairing scheme (not necessarily color superconductivity), as long as it
affects all quark flavors in similar way. In addition to the quark matter
and hadronic matter equation of state, we have also included a crustal EoS,
which is given by the Baym-Pethick-Sutherland EoS \cite{Baym1971}.

\begin{figure}
 \centering
 \vspace{1.0cm}
 \includegraphics[width=8.cm]{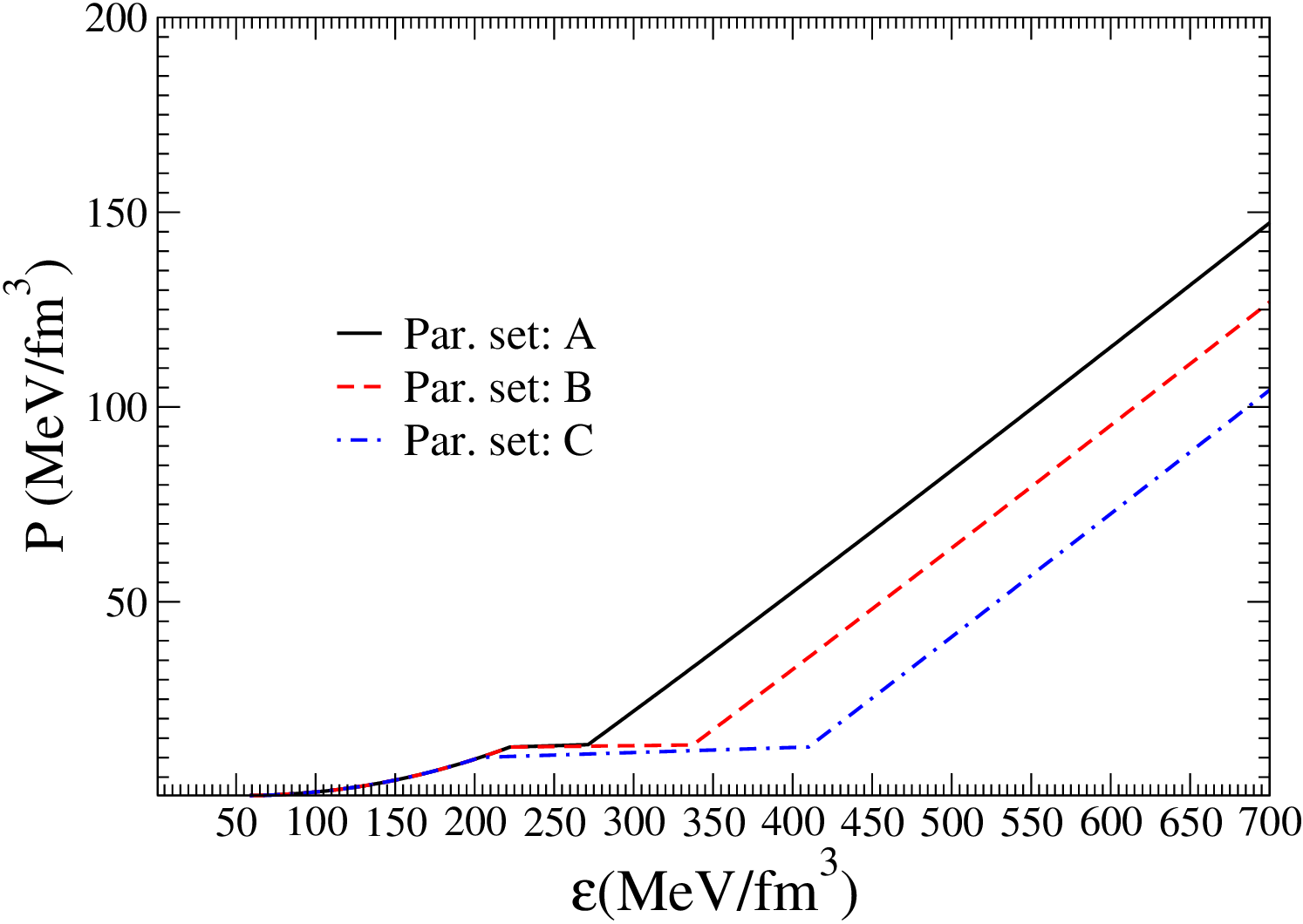}
 \caption{\label{eos}(Color online) Equation of state for the hadronic and quark
 phases shown for different parameter sets.}
 \end{figure}

The equation of state for the compact star is calculated assuming local
charge neutrality and beta equilibrium, as is usual for any bulk system. The
phase transition to quark matter is also calculated in the usual way, by
connecting the two EoS's at the point where the quark EoS has a higher
pressure. The different values used for the quark EoS parameters are
illustrated in Table \ref{table:Table_EOS}.  These parameters were chosen
to cover a wide range of quark core properties (mass, radius and composition),
which might potentially lead to different thermal evolution.
The final EoS's obtained are shown in Fig.~\ref{eos}, where we plot the pressure
as a function of energy density for the different EoS's investigated.
\begin{table}[t]
  \caption{\label{table:Table_EOS} Different parameter sets used to obtain the quark
matter EoS.}
\begin{ruledtabular}
\begin{tabular}{cccc}
 Parameter Set   & $B^{1/4}$~(MeV) & $m_s$~(MeV) & $\alpha_s$  \\
\hline
    & &  & \\
  A & 139  & 200   &  0.7     \\ 
  B & 150  & 175   &  0.4     \\ 
  C & 160  & 150   &  0.1     \\ 
\end{tabular}
\end{ruledtabular}
\end{table}

\begin{figure}
 \centering
 \vspace{1.0cm}
 \includegraphics[width=8.cm]{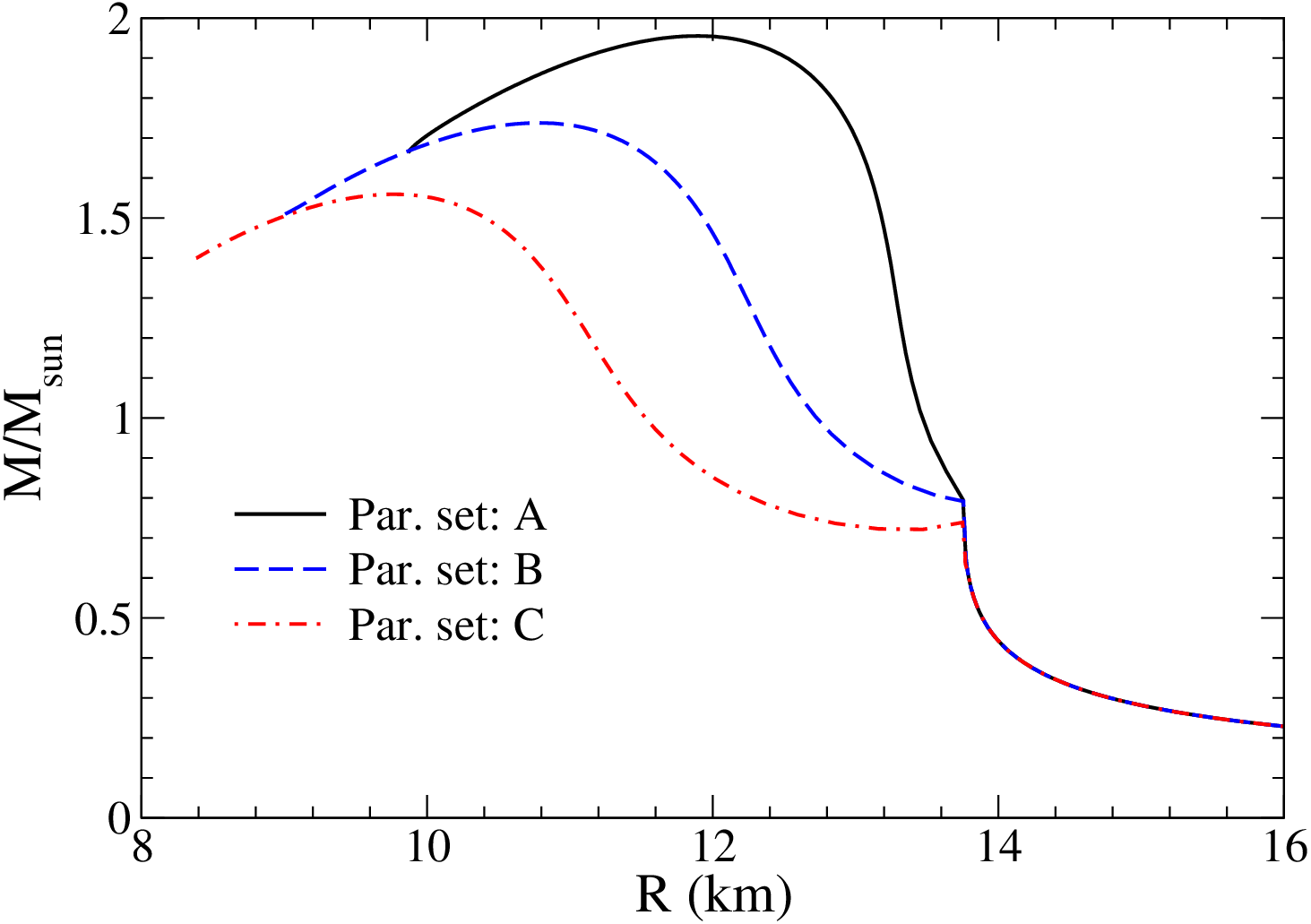}
 \caption{(Color online) Mass-radius diagram for stars whose
 equations of state are shown in Fig.~\ref{eos}. The sharp kinks denote the
 phase transition to quark matter.\label{mass}}
 \end{figure}

The mass-radius diagram for each EoS studied can be found in Fig.~\ref{mass}. 
The onset of quark matter can be easily identified as the sharp
kink found in the sequences. Note that for simplicity, we adopt a
Maxwell construction for the phase transition. We do not expect any qualitative
change of our conclusions if a more elaborate construction were used. 

\begin{figure}
 \centering
 \vspace{1.0cm}
 \includegraphics[width=8.cm]{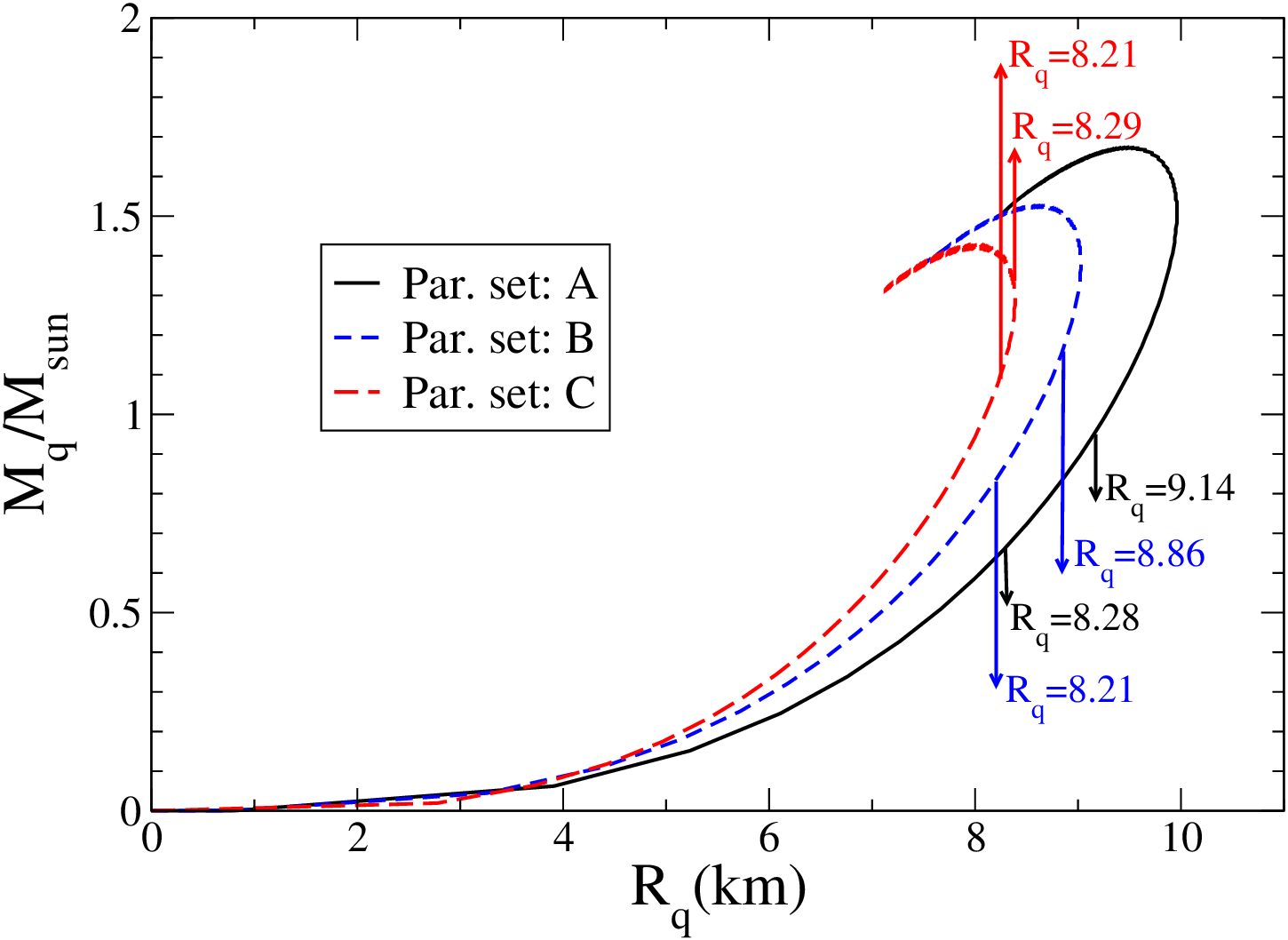}
 \caption{(Color online) Mass-radius diagram for the quark core of the
hybrid stars shown in  Fig.~\ref{mass}. Also shown the quark core radius for
the stars used in the cooling simulations (see table \ref{table:mass})
\label{Qmass}}
 \end{figure}

Given that each of the sequences shown in Fig.~\ref{mass} has a different quark matter
EoS, the quark core of the stars in each sequence should exhibit different properties. This is displayed in Fig.~\ref{Qmass} where the mass-radius diagram for the inner quark core of the hybrid stars is shown.

 Fig.~\ref{Qmass} clearly shows the distinct properties of the quark core for the
different
EoS's used. We now turn our attention to the thermal evolution of the stars shown in
Figs.~\ref{mass} and \ref{Qmass}, and determine the role that the different microscopic
compositions play in the thermal evolution of the object.

\section{Results} \label{Results}

The cooling of compact stars is given by the thermal balance and
thermal energy transport equation ($G = c = 1$) \cite{Weber}
\begin{eqnarray}
  \frac{ \partial (l e^{2\phi})}{\partial m}& = 
  &-\frac{1}{\rho \sqrt{1 - 2m/r}} \left( \epsilon_\nu 
    e^{2\phi} + c_v \frac{\partial (T e^\phi) }{\partial t} \right) \, , 
  \label{coeq1}  \\
  \frac{\partial (T e^\phi)}{\partial m} &=& - 
  \frac{(l e^{\phi})}{16 \pi^2 r^4 \kappa \rho \sqrt{1 - 2m/r}} 
  \label{coeq2} 
  \, .
\end{eqnarray}
In Eqs.~\ref{coeq1} and \ref{coeq2}, the structure of the star is given by the
variables  $r$, $\rho(r)$ and $m(r)$, that represent the radial distance, the
energy density, and the stellar mass, respectively. The thermal variables are
given by the temperature $T(r,t)$, luminosity $l(r,t)$, neutrino
emissivity $\epsilon_\nu(r,T)$, thermal conductivity
$\kappa(r,T)$ and specific heat $c_v(r,T)$.
The boundary conditions of (\ref{coeq1}) and (\ref{coeq2}) are determined
by the luminosity at the stellar center and at the surface. The
luminosity vanishes at the stellar center since at this point the heat flux is zero by definition. 
At the surface, the luminosity is defined by the relationship
between the mantle temperature and the temperature outside of the star
\cite{Gudmundsson1982,Gudmundsson1983,Page2006,Blaschke2000}. 
The microscopic inputs that enter Eqs.~(\ref{coeq1}) and (\ref{coeq2}) are
the neutrino emissivities, specific heat and thermal conductivity. For the
hadronic phase, the neutrino emission processes considered in this work are the
direct Urca, modified Urca and bremsstrahlung processes.  A
detailed review of the emissivities of such processes can be found in reference
\cite{Yakovlev2001a}.   We note that the neutrino emission involving neutrons, and the
neutron
specific heat are exponentially suppressed as the temperature drops below the neutron superfluidity
critical temperature. A description of the paring model, and
critical temperature can be found in reference \cite{Levenfish94}. As for the quark core,
we
consider the equivalent processes involving quarks. These are the quark direct Urca (QDU), quark
modified Urca (QMU), and quark bremsstrahlung processes (QBM). The emissivities of such
processes were calculated in \cite{IWAMOTO1982}. The specific heat of the
hadrons in the hadronic phase is given by the usual specific heat of fermions,
as described in \cite{Page2004}. For the quark phase we use the specific heat
as calculated in \cite{IWAMOTO1982}. Finally for the thermal conductivity of
the hadronic matter, we follow the calculations of \cite{Flowers1981}, and for
the quark matter we use the results of \cite{Haensel1991}.


In order to investigate  how the difference in the quark cores manifests itself in the
thermal evolution, we investigate the cooling of hybrid stars of same mass, but
different quark cores. The properties of such stars are listed in Table
\ref{table:star_props}.

\begin{table}[t]
  \caption{\label{table:star_props} Stellar properties of the stars used for the
thermal evolution investigation. All symbols have their
usual meaning, and $M_q$ and $R_q$ denotes the mass and radius of the quark matter core,
respectively. \label{table:mass}}
\begin{ruledtabular}
\begin{tabular}{ccccc}
 Parameter set & $M_q/M$  & $R_q$~ (km)  & $R$~ (km) &$M$~
($M_{\odot}$) \\
\hline
  A   & 0.48    & 8.28   &  13.26  & 1.33    \\
      & 0.60    & 9.14   &  13.15  & 1.55    \\ \hline
  B   & 0.62    & 8.21   &  12.20  & 1.33    \\
      & 0.74    & 8.86   &  11.82  & 1.55    \\ \hline
  C   & 0.82    & 8.21   &  10.89  & 1.33    \\ 
      & 0.89    & 8.29   &  10.03  & 1.55    \\ 
\end{tabular}
\end{ruledtabular}
\end{table}

\begin{figure}[h!]
 \centering
  \vspace{0.5 cm}
 \includegraphics[angle=0,width=8.7cm]{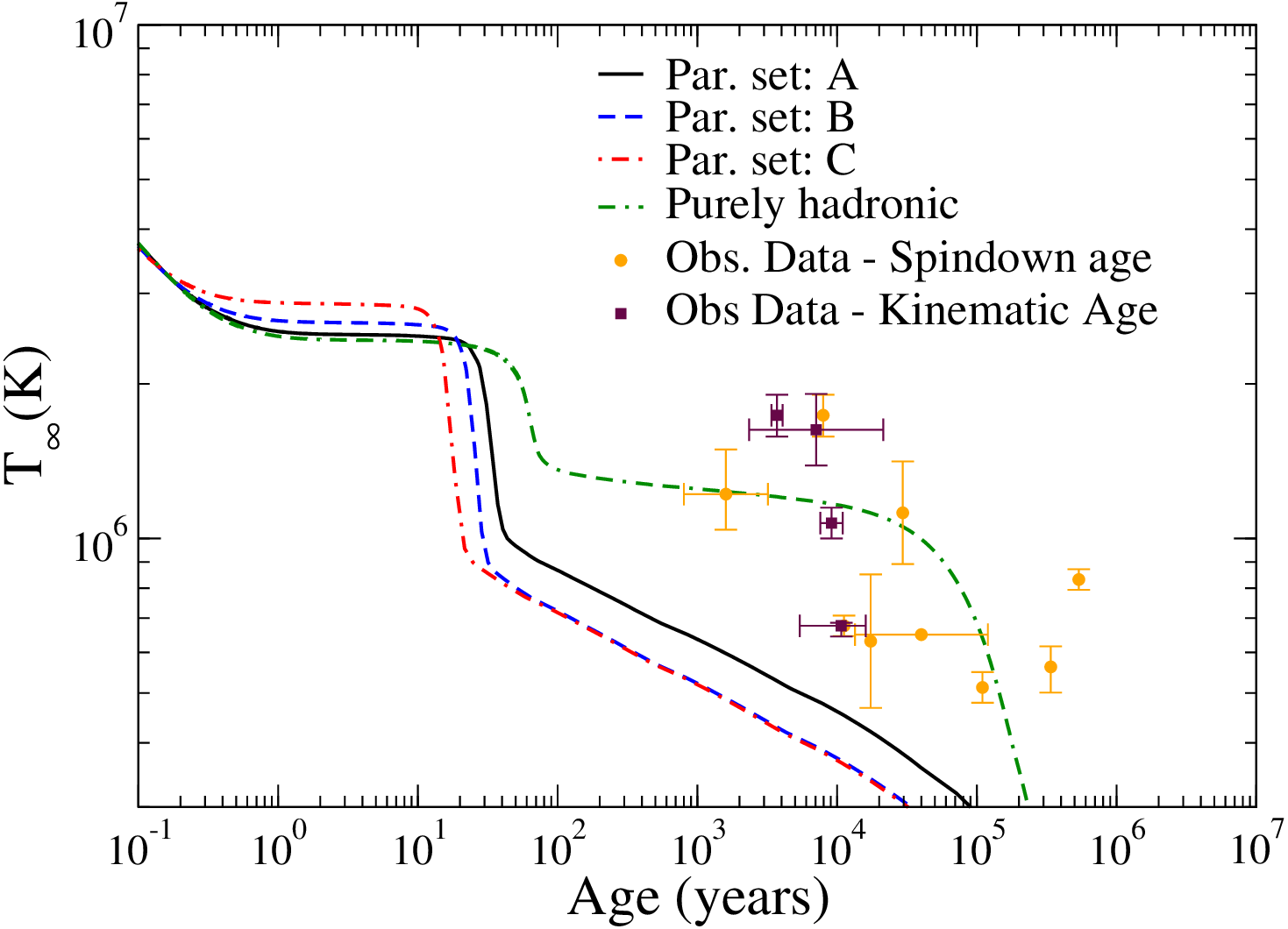}
 \caption{\label{cool_1}(Color online) Cooling of hybrid stars with gravitational mass of
1.33~$M_{\odot}$. $T_\infty$ denotes the temperature observed at infinity, and the x-axis the age in years. Also plotted is the cooling of the equivalent purely hadronic star with the same
mass. Two sets of observed  data were used, one whose age estimate is based on spin-down rate, and another that  is based on the kinematic age \cite{Page2004,Page2009}.}
 \end{figure}

\begin{figure}
 \centering
 \vspace{1.0cm}
 \includegraphics[angle=0,width=8.7cm]{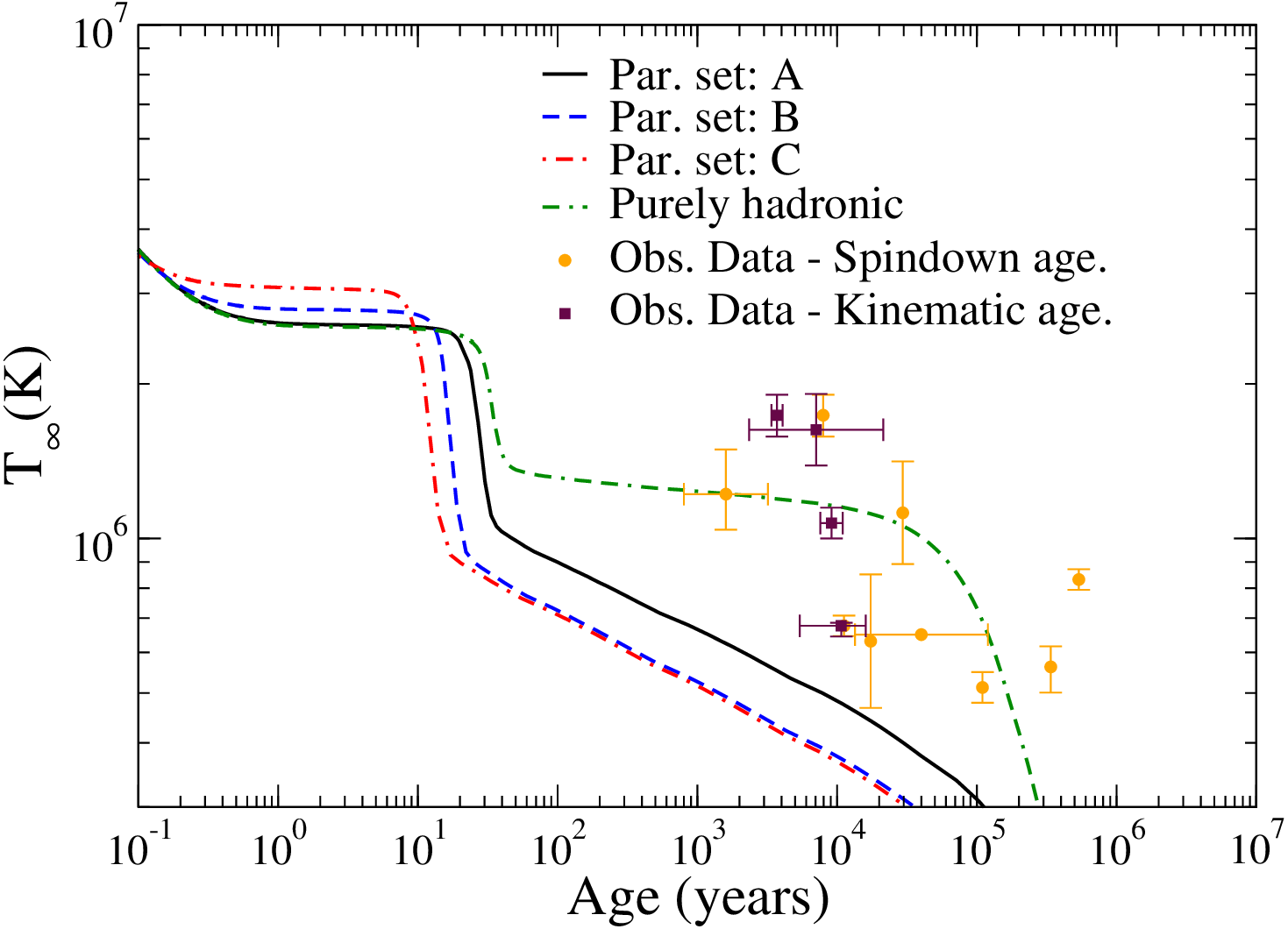}
 \caption{\label{cool_2}(Color online) Same as Figure~\ref{cool_1}, but for
 stellar mass of 1.55~$M_{\odot}$. }
 \end{figure}

The thermal evolution of the objects from table \ref{table:star_props} is illustrated in
Figs.~\ref{cool_1} and \ref{cool_2}, which shows the redshifted
temperature as a function of the age of the star. We have also
plotted prominent observed temperatures for a few pulsars. The data is
separated into two sets. The first set, labeled spin-down, contains the
temperature of objects whose age are estimated based on their spin-down age. The
second set, labeled kinematic, consists of age estimates based on the observed motion of the
object with respect to its progenitor supernova remnant. A detailed discussion about
the observed data can be found in references \cite{Page2004,Page2009}.

The results indicate that although there is
little difference between the cooling of stars with different gravitational masses, we can see
a significant difference between stars with different quark core properties. The
different cooling tracks exhibited by stars of different models (A, B and C) can
be attributed to the differences in the quark core. Figs.~\ref{cool_1} and
\ref{cool_2} show that for a more massive quark core, the cooling of the star
becomes faster. This is a result of the microscopic composition of such models,
that allow for more intense presence of the QDU (due to a higher electron
population). It is also important to understand that regardless of the total
mass of the star, the hadronic layer plays the same role. This is indicated by
the similar cooling behavior shown for the purely hadronic stars of masses 1.33
and 1.55~$M_{\odot}$ shown Figs.~\ref{cool_1} and \ref{cool_2}. The reason for
such behavior is the presence of the hadronic superfluidity, which suppresses
the hadronic DU process. Without superfluidity a neutron star with higher mass
could exhibit a faster cooling than a lower mass star, since the DU process
could be active in the former but not in the latter. The fact that the hadronic
layer plays the same role, regardless of the total mass of the object, is very
important for our analysis, since it allows us to better understand the effects
introduced by the different quark cores in the thermal evolution of the star.
 
\section{Role of Color Superconductivity} \label{SC}

It seems clear that the presence of the quark core enhances the cooling of the
star to a point that is it in clear disagreement with the observed data, whereas
the purely hadronic model seems to do a (slightly) better job. We note however
that results presented so far do not represent the whole picture. As mentioned
in section \ref{Model}, strange quark matter is expected to be in a
superconducting phase. The most likely condensation pattern for high densities
is the Color-Flavor-Locked phase, in which all quarks are paired. Because of 
pairing the direct Urca process is suppressed by a factor $e^{-\Delta/T}$,
and the modified Urca and the Bremsstrahlung process by a factor
$e^{-2\Delta/T}$, for $T \leq T_c$, where $\Delta$ is the gap parameter
for the CFL phase and $T_c$ is the critical temperature below which strange
matter undergoes a
phase transition into CFL matter. In addition to changes in the neutrino
emissivities, the specific heat of quark matter is also modified, being reduced
by the factor $3.2 (T_c/T)*(2.5 - 1.7(T/T_c)
+3.6(T/T_c)^2)e^{-\Delta/T}$ \cite{Blaschke2000}. The critical temperature
for the CFL phase is currently not known, although it is believed to be smaller
than the standard Bardeen-Cooper-Schrieffer ($T_c \simeq 0.57 \Delta$), due
to instanton$-$anti-instanton effects. In this work we follow the footsteps of
the authors in \cite{Blaschke2000}, and use $T_c = 0.4\Delta$.

 In addition to the traditional neutrino emissivities, one
could also include the emissivities from massive and massless Goldstone modes.
These emissions, as
discussed in \cite{Jaikumar2002}, are small, and should have small effects on
the thermal evolution of the object if more powerful emission processes are
present.
In addition to that, we have also not considered the contributions of photons,
and Goldstone modes
to the thermal conductivity of CFL quark matter \cite{Shovkovy2002}. If these
modes are
considered, the thermal conductivity of quark matter should be higher than that
calculated in
\cite{Haensel1991}. Not denying that such contributions are important, we do not
expect that they
would strongly affect our results. As discussed in \cite{Shovkovy2002},
a higher thermal conductivity would only mean that the quark core will be in
thermal equilibrium with the hadronic phase more quickly. We intend to include
these contributions in future
calculations.

\begin{figure}
 \centering
 \vspace{1.0cm}
 \includegraphics[angle=0,width=8.7cm]{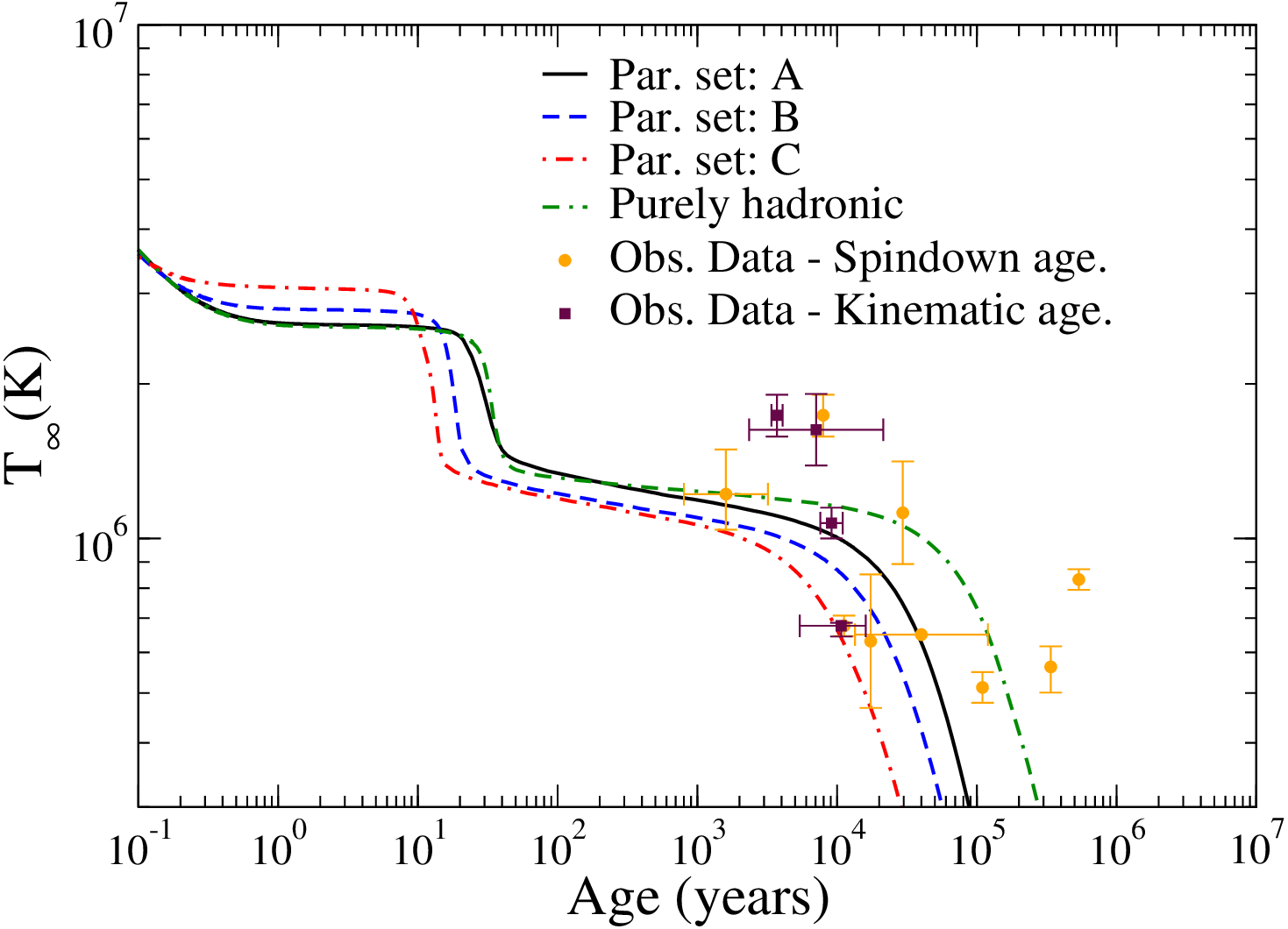}
 \caption{\label{cool_3}(Color online)  Cooling of hybrid stars with gravitational mass of
1.55~$M_{\odot}$, and $\Delta = 0.1$ MeV. Axis and observed data are the same as in
Fig.~\ref{cool_1} }
 \end{figure}

\begin{figure}
 \centering
 \vspace{1.0cm}
 \includegraphics[angle=0,width=8.7cm]{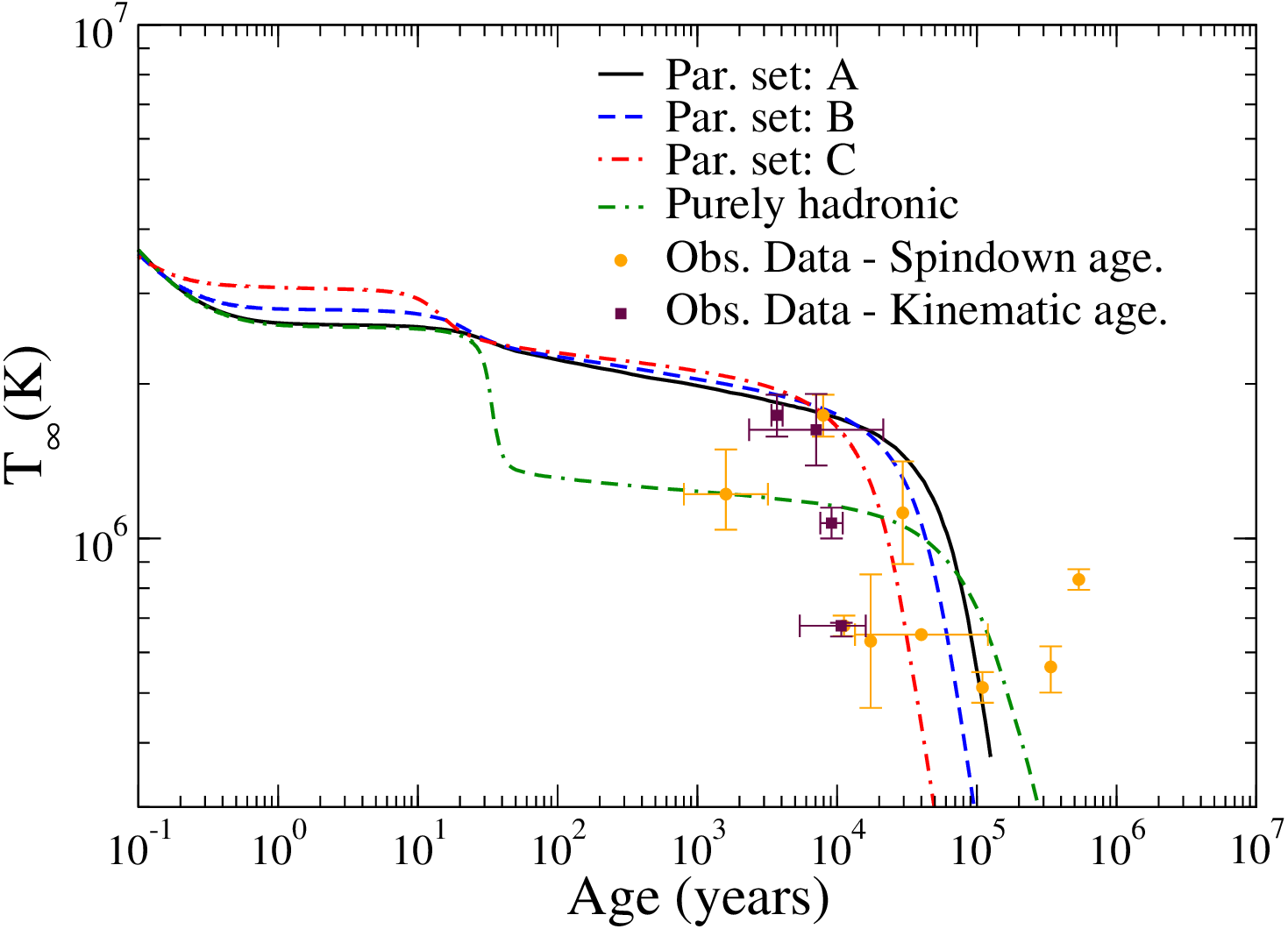}
 \caption{\label{cool_4}(Color online) Same as Fig. \ref{cool_4} but for
 $\Delta = 1.0$ MeV MeV.}
 \end{figure}

We now show in Figs.~\ref{cool_3} and \ref{cool_4}  the results for the
cooling of hybrid stars of 1.55~$M_{\odot}$, whose quark core is composed of strange quark matter in
the CFL phase with $\Delta = 0.1$ MeV and $\Delta = 1.0$ MeV, respectively.

As discussed in the last section, the hadronic layer plays a small role in the
cooling of hybrid stars, with the quark core dominating the thermal evolution (at least
for the first 1000 years). Following this logic it is only obvious that the
introduction of the CFL
phase should alter the cooling substantially. This is indeed the case, as shown in
Figs.~\ref{cool_3} and \ref{cool_4}. We see
that objects with $\Delta = 0.1$ MeV exhibit a much slower cooling than their unpaired
counterparts ( Figs.~\ref{cool_1} and \ref{cool_2}). The slower cooling is due the suppression of
the neutrino emission processes in the quark core resulting from quark pairing.
Fig.~\ref{cool_4} shows that a higher value of $\Delta$ will result in even
slower cooling as  a higher $\Delta$ will lead to stronger suppression. We note, however, that for
$\Delta \geq 1.0$ the resulting thermal evolution is almost
the same as that of $\Delta = 1.0$. This is due to the fact that the exponential
$\exp{(-\Delta/T)}$ effectively saturates for $\Delta > 1.0$ MeV. 

\begin{figure}
 \centering
 \vspace{1.0cm}
 \includegraphics[angle=0,width=8.7cm]{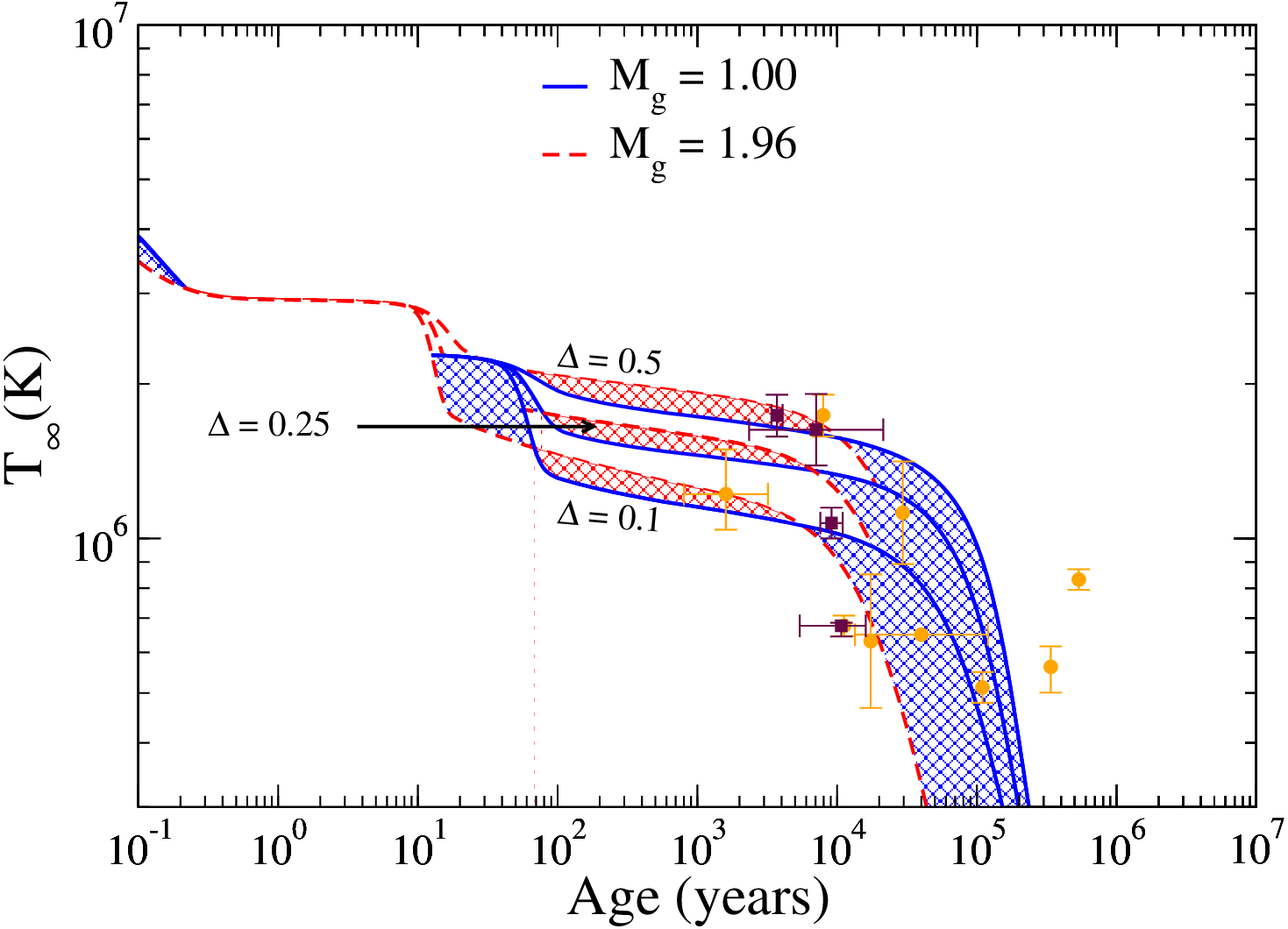}
 \caption{\label{cool_DELTA1}(Color online) Cooling of hybrid stars (model A),
for different values
of the CFL gap ($\Delta$).}
 \end{figure}

\begin{figure}
 \centering
 \vspace{1.0cm}
 \includegraphics[angle=0,width=8.7cm]{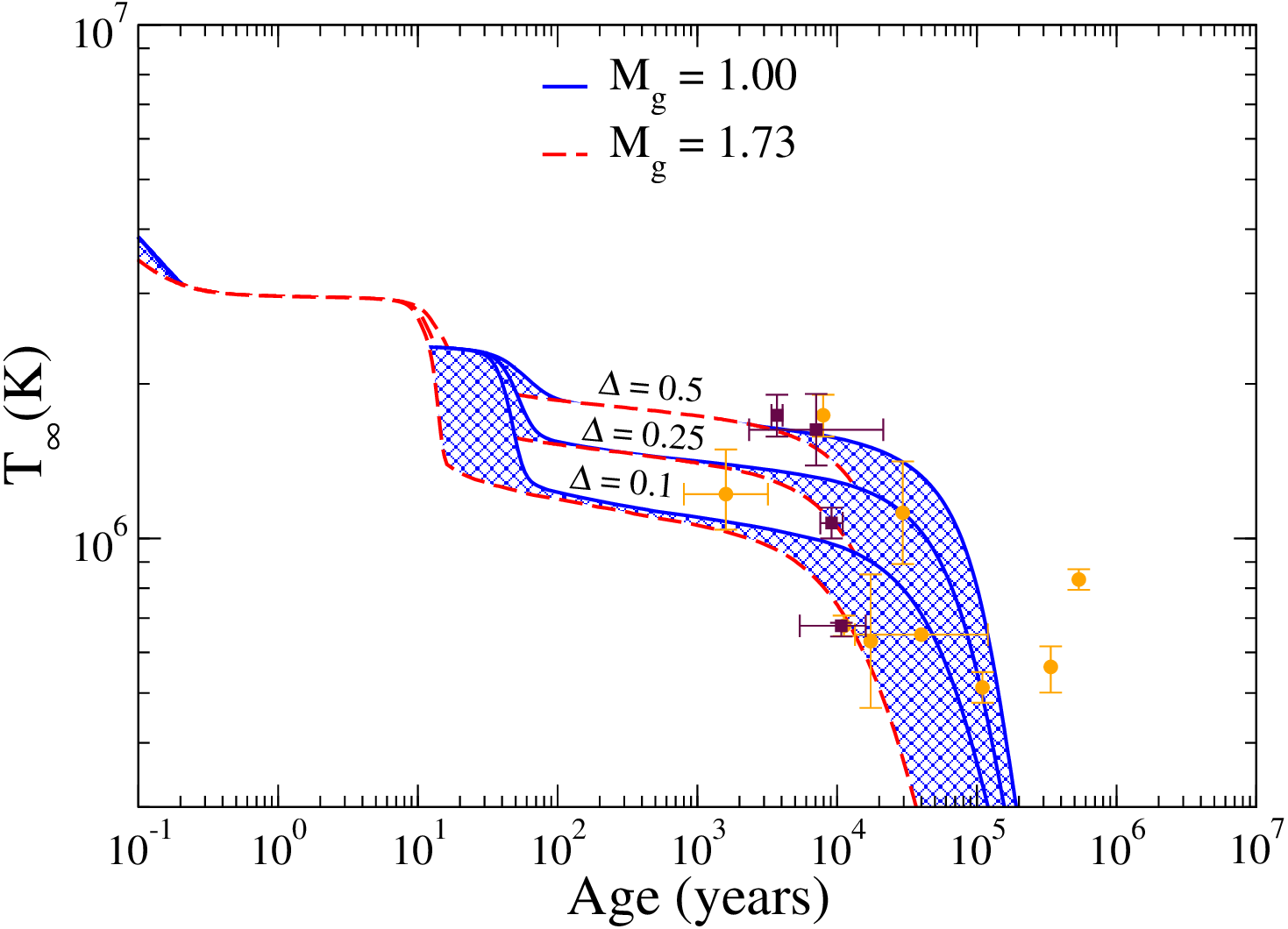}
 \caption{\label{cool_DELTA2}(Color online) Cooling of hybrid stars (model B),
for different values
of the CFL gap ($\Delta$).}
 \end{figure}

\begin{figure}
 \centering
 \vspace{1.0cm}
 \includegraphics[angle=0,width=8.7cm]{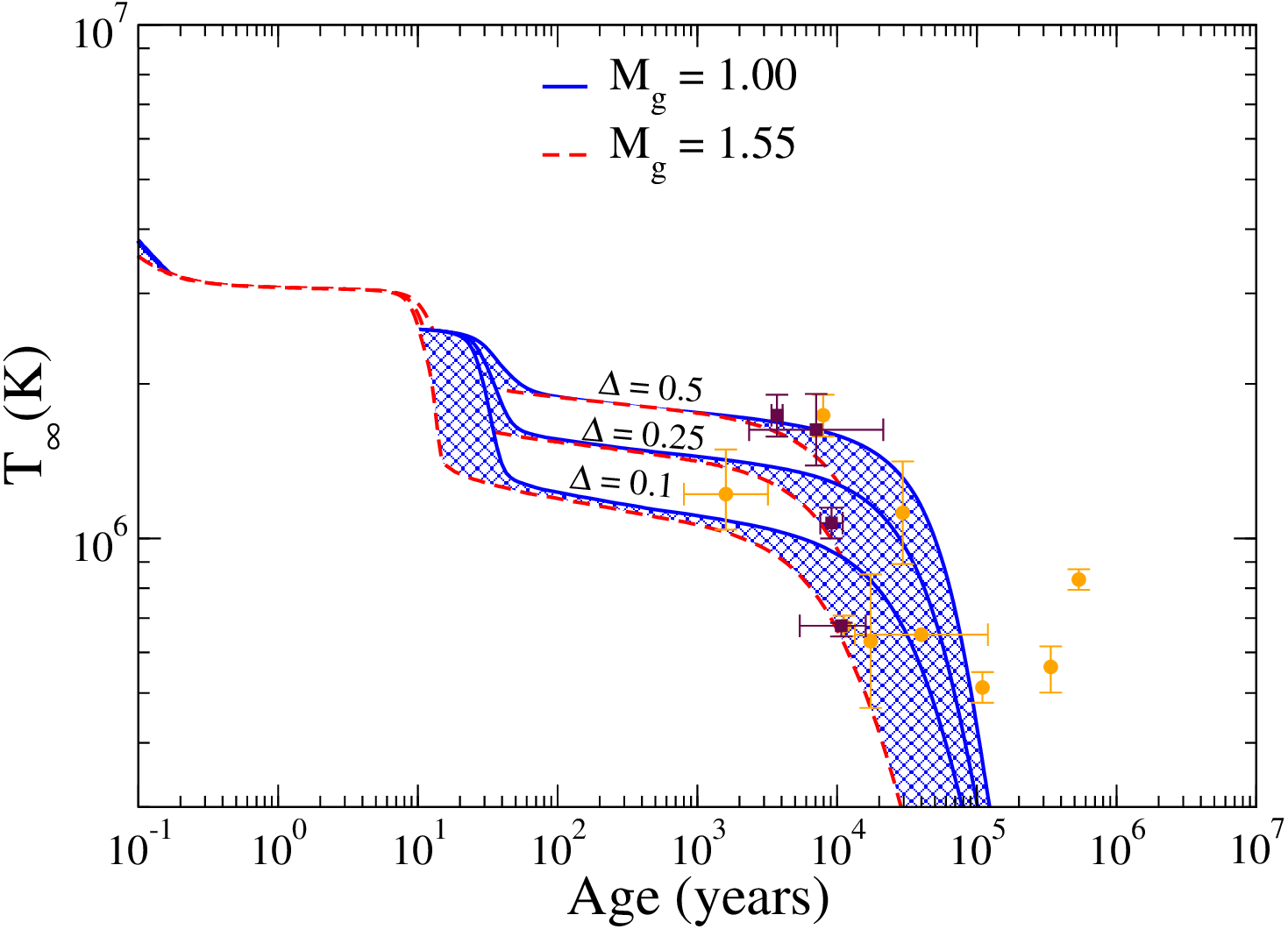}
 \caption{\label{cool_DELTA3}(Color online) Cooling of hybrid stars (model C),
for different values
of the CFL gap ($\Delta$).}
 \end{figure}

Figs.~\ref{cool_3} and \ref{cool_4} also show that hybrid stars, whose quark
cores are in a CFL
phase, agree relatively well with the observed data. We see, however, that smaller values of
$\Delta$ ($\sim 0.1$ MeV) lead to agreement with only a few objects (those with
lower temperatures), while $\Delta \sim 1.0$ MeV leads to agreement with those
with higher temperature. This behavior hints that the best agreement will
possibly be obtained for intermediate values of $\Delta$, if one means to
interpret the whole set of observed data as hybrid stars.
In order to investigate this we explore different values of $\Delta$, and 
calculate the cooling band spanned by the cooling curves of stars with 1.00 
$M_{\odot}$ and the maximum mass of each respective parameter set. This result
is shown in Figs.~\ref{cool_DELTA1}-\ref{cool_DELTA3}.


 The results shown in Figs.~\ref{cool_DELTA1}-\ref{cool_DELTA3} indicate
that the observed data can be better explained by hybrid stars whose quark phase
is in a CFL state with pairing gaps between 0.25 and 0.5 MeV. These results also
show us that the difference between models B and C is very subtle, with both
these models describing the same data points. Model A, however, is
substantially different,as can be seen in Fig.~\ref{cool_DELTA1}. One can see
that as opposed to B and C, in the range of age between $10^2$ to $10^4$
years, the most massive star in Model A exhibits a slower cooling. This
behaviour is due to less massive quark cores in this model. We note
that
there are still two objects that do not fall within the range of the cooling
curves, these are the two oldest stars in the graph. We believe that this might
be due to uncertainties in the estimate of their spin-down age. As pointed out
in \cite{Espinoza2011}, the spin-down age may be an overestimation of their
actual age. If indeed that is the case these objects may be younger than they
appear, in which case they may fall into the range defined by the cooling
curves. The analysis of the age estimate of these objects, however, is beyond
the
scope of this paper.

\section{Discussion \& Conclusions} \label{conc}

It is the purpose of this paper to investigate the importance of the   quark core
properties to the cooling of hybrid stars. Although there have been a
handful of papers analyzing different models for hybrid stars, and their agreement with
observational mass and radius
constraints \cite{Grigorian2005,Blaschke2006,Negreiros2010,Dexheimer2010,
Kurkela2010,Blaschke2010}, we intend to provide a complementary study, in which
observed temperatures of pulsars, together with cooling simulations of hybrid stars are used to
infer the importance of the microscopic properties of the quark core
to the thermal evolution.
To achieve that goal, we have used a MIT bag model to describe the quark phase, and a relativistic
mean field model for the hadron phase. Different values for  microscopic parameters of the
quark phase were used, which led to quark cores with different properties.

 The results obtained indicate that it is possible to explain the observed
thermal properties of pulsars with a hybrid star model. Furthermore,  the
best agreement with the observed data was obtained for hybrid stars with a
CFL quark core with $\alpha_s=0.4$, $B^{1/4} = 150$ MeV, $m_s = 175$ MeV and with $\Delta
= 0.25-1.0$ MeV. We note that parameter set A ($\alpha_s=0.7$, $B^{1/4} =
139$ MeV, $m_s = 200$) was also in good agreement with
the observed data. Furthermore, this parameter set has the advantage of
agreeing with the observed mass of J1614-2230 (1.96~$\pm$ 0.04M$_\odot$)
\cite{Demorest2011}. Parameter set A assumes, however, that $\alpha_s = 0.7$
which might be considered too high for the first order corrections used in the
calculation of the quark matter EoS. Because of this, the results obtained
from parameter set A should be regarded with care, which is the reason why we
selected parameter set B (which also exhibits satisfactory agreement with
cooling data, albeit not with the mass of J1614-2230) as that with the best
agreement with the observed cooling data (given its more reliable microscopic
model). We can
however, learn something from the results of parameter set A, i.e. that the
cooling will present a better agreement with the observed data for models that
yield less massive quark cores, and lower electron population (as in parameter
set A).  For higher values of the gap parameter ($\Delta
\geq 10$ MeV), the band of possible cooling curves becomes too narrow, which worsens the agreement with the observed data. We note that this result is independent of the equation of state used for the hadronic matter, as long as the latter presents a suppression of fast neutrino emission processes. This is very reasonable, considering that the hadronic layer is constrained to the lower density regime, in which neutron superfluidity is very likely to take place. 

It is important to remember that in this paper we have considered a somewhat simple microscopic model, and a more sophisticated study, considering for example different degrees of freedom, and chiral symmetry restoration \cite{Negreiros2010,Dexheimer2010,Blaschke2010}
would certainly be of interest. Work in that direction is currently in progress.
We believe, however, that even for a more sophisticated microscopic model our
conclusions should still hold.  If so, this would imply that a quark
core with the properties similar to those of the best case scenario (see
parameters above) should provide the best agreement with the observed data.

We note that, although we have considered neutron superfluidity, we have not
included proton superconductivity in the core, nor the emission of neutrinos due to the pair
breaking/formation process. The aforementioned processes were the subject of recent papers
\cite{Page2011a,Yakovlev2011a} aimed at modeling the observed temperature evolution of the
neutron star in Cas A. The inclusion of proton superconductivity is still
speculative, and since we do not have a clear picture of how (or if) it 
manifests itself in a high-density environment (see ref. \cite{Page2011b} and
references therein), we chose to take a conservative approach and only consider
neutron superfluidity. One could also include the pair breaking/formation
process, however, this should have little effect on our results. Differently
from references \cite{Page2011a,Yakovlev2011a}, we allow fast neutrino processes
to take place (hence the dip in the cooling curve at around 100 years), which
(even suppressed by pairing) dominate over the other neutrino emission
processes. Therefore, by allowing fast emission processes in the core, we
cannot describe the observed behaviour of the neutron star in Cas A. We note
that the cooling of high mass stars of parameter set A (in a CFL phase) present
temperatures and ages comparable to those of Cas A, but the slope of the
temperature evolution is wrong. This indicates that either the neutron star in
Cas A has a mass smaller than the direct Urca threshold (and thus has no fast
neutrino emission processes), or that our model needs further
refinements. We should also point out that the authors of
\cite{Noda2011} have shown that a hybrid star model can be used to model the
observed behavior of Cas A, indicating that the recent observations do
not rule out hybrid stars. Furthermore, it has been recently proposed that the
behavior of Cas A can be explained as the late onset of the Direct Urca process
in spinning down neutron stars \cite{Negreiros2011}. We note that the same
principle could be applied to hybrid stars, where the transition to quark matter
could be delayed due to spin-down. This study will be the object of a future
investigation.

 The purpose of this paper was to determine if quark cores with different properties would
lead to different thermal evolution, in a way that one could use this
information to constrain the microscopic properties of quark matter in hybrid
stars. We have found a set of parameters that leads to a reasonably good
agreement with the observed data of cooling neutron stars, although not
all of them agree with the observed mass of J1614-2230.
It is evident that there is room for improvement in the approach we used to
model quark matter microscopically, and it is our intent to use the results and
constraints put forth in this work, in addition to the latest observational
data, to refine the current models for the EoS of quark matter. 
\\


\begin{thebibliography}{10}%
\makeatletter
\providecommand \@ifxundefined [1]{%
 \ifx #1\undefined \expandafter \@firstoftwo
 \else \expandafter \@secondoftwo
\fi
}%
\providecommand \@ifnum [1]{%
 \ifnum #1\expandafter \@firstoftwo
 \else \expandafter \@secondoftwo
\fi
}%
\providecommand \enquote [1]{``#1''}%
\providecommand \bibnamefont  [1]{#1}%
\providecommand \bibfnamefont [1]{#1}%
\providecommand \citenamefont [1]{#1}%
\providecommand\href[0]{\@sanitize\@href}%
\providecommand\@href[1]{\endgroup\@@startlink{#1}\endgroup\@@href}%
\providecommand\@@href[1]{#1\@@endlink}%
\providecommand \@sanitize [0]{\begingroup\catcode`\&12\catcode`\#12\relax}%
\@ifxundefined \pdfoutput {\@firstoftwo}{%
 \@ifnum{\z@=\pdfoutput}{\@firstoftwo}{\@secondoftwo}%
}{%
 \providecommand\@@startlink[1]{\leavevmode}%
 \providecommand\@@endlink[0]{}%
}{%
 \providecommand\@@startlink[1]{%
  \leavevmode
  \pdfstartlink
   attr{/Border[0 0 1 ]/H/I/C[0 1 1]}%
   user{/Subtype/Link/A<</Type/Action/S/URI/URI(#1)>>}%
  \relax
 }%
 \providecommand\@@endlink[0]{\pdfendlink}%
}%
\providecommand \url  [0]{\begingroup\@sanitize \@url }%
\providecommand \@url [1]{\endgroup\@href {#1}{\urlprefix}}%
\providecommand \urlprefix [0]{URL }%
\providecommand \Eprint[0]{\href }%
\@ifxundefined \urlstyle {%
  \providecommand \doi [1]{doi:\discretionary{}{}{}#1}%
}{%
  \providecommand \doi [0]{doi:\discretionary{}{}{}\begingroup
  \urlstyle{rm}\Url }%
}%
\providecommand \doibase [0]{http://dx.doi.org/}%
\providecommand \Doi[1]{\href{\doibase#1}}%
\providecommand \bibAnnote [3]{%
  \BibitemShut{#1}%
  \begin{quotation}\noindent
    \textsc{Key:}\ #2\\\textsc{Annotation:}\ #3%
  \end{quotation}%
}%
\providecommand \bibAnnoteFile [2]{%
  \IfFileExists{#2}{\bibAnnote {#1} {#2} {\input{#2}}}{}%
}%
\providecommand \typeout [0]{\immediate \write \m@ne }%
\providecommand \selectlanguage [0]{\@gobble}%
\providecommand \bibinfo [0]{\@secondoftwo}%
\providecommand \bibfield [0]{\@secondoftwo}%
\providecommand \translation [1]{[#1]}%
\providecommand \BibitemOpen[0]{}%
\providecommand \bibitemStop [0]{}%
\providecommand \bibitemNoStop [0]{.\EOS\space}%
\providecommand \EOS [0]{\spacefactor3000\relax}%
\providecommand \BibitemShut [1]{\csname bibitem#1\endcsname}%
\bibitem{Weber}%
  \BibitemOpen
  \bibfield{author}{%
  \bibinfo {author} {\bibfnamefont{F.}~\bibnamefont{Weber}},\ }%
  \emph{\bibinfo {title} {{Pulsars as astrophysical laboratories for nuclear
  and particle physics}}},\ \bibinfo {edition} {1st}\ ed.\ (\bibinfo
  {publisher} {Institute of Physics},\ \bibinfo {address} {Bristol},\ \bibinfo
  {year} {1999})%
  \bibAnnoteFile{NoStop}{Weber}%
\bibitem{Glendenning2000}%
  \BibitemOpen
  \bibfield{author}{%
  \bibinfo {author} {\bibfnamefont{N.~K.}\ \bibnamefont{Glendenning}},\ }%
  \emph{\bibinfo {title} {{Compact stars : nuclear physics, particle physics,
  and general relativity}}},\ \bibinfo {edition} {1st}\ ed.\ (\bibinfo
  {publisher} {Springer},\ \bibinfo {year} {2000})%
  \bibAnnoteFile{NoStop}{Glendenning2000}%
\bibitem{Ozel2010}%
  \BibitemOpen
  \bibfield{author}{%
  \bibinfo {author} {\bibfnamefont{F.}~\bibnamefont{Ozel}}, \bibinfo {author}
  {\bibfnamefont{G.}~\bibnamefont{Baym}},\ and\ \bibinfo {author}
  {\bibfnamefont{T.}~\bibnamefont{Guver}},\ }%
  \bibfield{journal}{%
  \bibinfo {journal} {arXiv/1002.3153}}%
   (\bibinfo {year} {2010})%
  \bibAnnoteFile{NoStop}{Ozel2010}%
\bibitem{Steiner2010}%
  \BibitemOpen
  \bibfield{author}{%
  \bibinfo {author} {\bibfnamefont{A.~W.}\ \bibnamefont{Steiner}}, \bibinfo
  {author} {\bibfnamefont{J.}~\bibnamefont{Lattimer}},\ and\ \bibinfo {author}
  {\bibfnamefont{E.~F.}\ \bibnamefont{Brown}},\ }%
  \bibfield{journal}{%
  \bibinfo {journal} {arXiv/1005.0811}}%
   (\bibinfo {year} {2010})%
  \bibAnnoteFile{NoStop}{Steiner2010}%
\bibitem{Page2004}%
  \BibitemOpen
  \bibfield{author}{%
  \bibinfo {author} {\bibfnamefont{D.}~\bibnamefont{Page}}, \bibinfo {author}
  {\bibfnamefont{J.}~\bibnamefont{Lattimer}}, \bibinfo {author}
  {\bibfnamefont{M.}~\bibnamefont{Prakash}},\ and\ \bibinfo {author}
  {\bibfnamefont{A.~W.}\ \bibnamefont{Steiner}},\ }%
  \bibfield{journal}{%
  \bibinfo {journal} {The Astrophysical Journal Supplement Series}\ }%
  \textbf{\bibinfo {volume} {155}},\ \bibinfo {pages} {623} (\bibinfo {year}
  {2004})%
  \bibAnnoteFile{NoStop}{Page2004}%
\bibitem{Page2009}%
  \BibitemOpen
  \bibfield{author}{%
  \bibinfo {author} {\bibfnamefont{D.}~\bibnamefont{Page}}, \bibinfo {author}
  {\bibfnamefont{J.}~\bibnamefont{Lattimer}}, \bibinfo {author}
  {\bibfnamefont{M.}~\bibnamefont{Prakash}},\ and\ \bibinfo {author}
  {\bibfnamefont{A.~W.}\ \bibnamefont{Steiner}},\ }%
  \bibfield{journal}{%
  \bibinfo {journal} {The Astrophysical Journal}\ }%
  \textbf{\bibinfo {volume} {707}},\ \bibinfo {pages} {1131} (\bibinfo {year}
  {2009})%
  \bibAnnoteFile{NoStop}{Page2009}%
\bibitem{Page2006}%
  \BibitemOpen
  \bibfield{author}{%
  \bibinfo {author} {\bibfnamefont{D.}~\bibnamefont{Page}}, \bibinfo {author}
  {\bibfnamefont{U.}~\bibnamefont{Geppert}},\ and\ \bibinfo {author}
  {\bibfnamefont{F.}~\bibnamefont{Weber}},\ }%
  \bibfield{journal}{%
  \bibinfo {journal} {Nuclear Physics A}\ }%
  \textbf{\bibinfo {volume} {777}},\ \bibinfo {pages} {497} (\bibinfo {year}
  {2006})%
  \bibAnnoteFile{NoStop}{Page2006}%
\bibitem{Lattimer1991}%
  \BibitemOpen
  \bibfield{author}{%
  \bibinfo {author} {\bibfnamefont{J.}~\bibnamefont{Lattimer}}, \bibinfo
  {author} {\bibfnamefont{C.}~\bibnamefont{Pethick}}, \bibinfo {author}
  {\bibfnamefont{M.}~\bibnamefont{Prakash}},\ and\ \bibinfo {author}
  {\bibfnamefont{P.}~\bibnamefont{Haensel}},\ }%
  \bibfield{journal}{%
  \bibinfo {journal} {Physical review letters}\ }%
  \textbf{\bibinfo {volume} {66}},\ \bibinfo {pages} {27012704} (\bibinfo
  {year} {1991})%
  \bibAnnoteFile{NoStop}{Lattimer1991}%
\bibitem{Blaschke2000}%
  \BibitemOpen
  \bibfield{author}{%
  \bibinfo {author} {\bibfnamefont{D.}~\bibnamefont{Blaschke}}, \bibinfo
  {author} {\bibfnamefont{T.}~\bibnamefont{Klahn}},\ and\ \bibinfo {author}
  {\bibfnamefont{D.}~\bibnamefont{Voskresensky}},\ }%
  \bibfield{journal}{%
  \bibinfo {journal} {The Astrophysical Journal}\ }%
  \textbf{\bibinfo {volume} {533}},\ \bibinfo {pages} {406412} (\bibinfo {year}
  {2000})%
  
 \bibitem{Schaab94}
C Schaab,\ F. Weber,\ M. Weigel,\ \& N. K. Glendenning, Nuclear Phys A {\bf 605}, 531 (1996). 
  
\bibitem{Levenfish94}
K. P. Levenfish,\ \& D. G. Yakovlev,\ Astronomy Letters {\bf 20},\ 43 (1994).  

\bibitem{Page2011b}
D. Page,\ M. Prakash,\ J. M. Lattimer,\ \& A. W. Steiner, astro-ph/1110.5116 (2011).
  
  \bibAnnoteFile{NoStop}{Blaschke2000}%
\bibitem{Grigorian2005}%
  \BibitemOpen
  \bibfield{author}{%
  \bibinfo {author} {\bibfnamefont{H.}~\bibnamefont{Grigorian}}, \bibinfo
  {author} {\bibfnamefont{D.}~\bibnamefont{Blaschke}},\ and\ \bibinfo {author}
  {\bibfnamefont{D.}~\bibnamefont{Voskresensky}},\ }%
  \bibfield{journal}{%
  \bibinfo {journal} {Physical Review C}\ }%
  \textbf{\bibinfo {volume} {71}},\ \bibinfo {pages} {1} (\bibinfo {year}
  {2005})%
  \bibAnnoteFile{NoStop}{Grigorian2005}%
\bibitem{Blaschke2006}%
  \BibitemOpen
  \bibfield{author}{%
  \bibinfo {author} {\bibfnamefont{D.}~\bibnamefont{Blaschke}}, \bibinfo
  {author} {\bibfnamefont{D.}~\bibnamefont{Voskresensky}},\ and\ \bibinfo
  {author} {\bibfnamefont{H.}~\bibnamefont{Grigorian}},\ }%
  \bibfield{journal}{%
  \bibinfo {journal} {Nuclear Physics A}\ }%
  \textbf{\bibinfo {volume} {774}},\ \bibinfo {pages} {815} (\bibinfo {year}
  {2006})%
  \bibAnnoteFile{NoStop}{Blaschke2006}%
\bibitem{Jaikumar2002}%
  \BibitemOpen
  \bibfield{author}{%
  \bibinfo {author} {\bibfnamefont{P.}~\bibnamefont{Jaikumar}}, \bibinfo
  {author} {\bibfnamefont{M.}~\bibnamefont{Prakash}},\ and\ \bibinfo {author}
  {\bibfnamefont{T.}~\bibnamefont{Sch\"{a}fer}},\ }%
  \bibfield{journal}{%
  \bibinfo {journal} {Physical Review D}\ }%
  \textbf{\bibinfo {volume} {66}},\ \bibinfo {pages} {1} (\bibinfo {year}
  {2002})%
  \bibAnnoteFile{NoStop}{Jaikumar2002}%
\bibitem{Shovkovy2002}%
  \BibitemOpen
  \bibfield{author}{%
  \bibinfo {author} {\bibfnamefont{I.}~\bibnamefont{Shovkovy}}\ and\ \bibinfo
  {author} {\bibfnamefont{P.}~\bibnamefont{Ellis}},\ }%
  \bibfield{journal}{%
  \bibinfo {journal} {Physical Review C}\ }%
  \textbf{\bibinfo {volume} {66}},\ \bibinfo {pages} {1} (\bibinfo {year}
  {2002})%
  \bibAnnoteFile{NoStop}{Shovkovy2002}%
\bibitem{Tolman1939}%
  \BibitemOpen
  \bibfield{author}{%
  \bibinfo {author} {\bibfnamefont{R.}~\bibnamefont{Tolman}},\ }%
  \bibfield{journal}{%
  \bibinfo {journal} {Physical Review}\ }%
  \textbf{\bibinfo {volume} {55}},\ \bibinfo {pages} {364} (\bibinfo {year}
  {1939})%
  \bibAnnoteFile{NoStop}{Tolman1939}%
\bibitem{Oppenheimer1939}%
  \BibitemOpen
  \bibfield{author}{%
  \bibinfo {author} {\bibfnamefont{J.}~\bibnamefont{Oppenheimer}}\ and\
  \bibinfo {author} {\bibfnamefont{G.}~\bibnamefont{Volkoff}},\ }%
  \bibfield{journal}{%
  \bibinfo {journal} {Physical Review}\ }%
  \textbf{\bibinfo {volume} {55}},\ \bibinfo {pages} {374} (\bibinfo {year}
  {1939})%
  \bibAnnoteFile{NoStop}{Oppenheimer1939}%
\bibitem{Grigorian2004}%
  \BibitemOpen
  \bibfield{author}{%
  \bibinfo {author} {\bibfnamefont{H.}~\bibnamefont{Grigorian}}, \bibinfo
  {author} {\bibfnamefont{D.}~\bibnamefont{Blaschke}},\ and\ \bibinfo {author}
  {\bibfnamefont{D.}~\bibnamefont{Aguilera}},\ }%
  \bibfield{journal}{%
  \bibinfo {journal} {Physical Review C}\ }%
  \textbf{\bibinfo {volume} {69}},\ \bibinfo {pages} {1} (\bibinfo {year}
  {2004})%
  \bibAnnoteFile{NoStop}{Grigorian2004}%
\bibitem{Negreiros2010}%
  \BibitemOpen
  \bibfield{author}{%
  \bibinfo {author} {\bibfnamefont{R.}~\bibnamefont{Negreiros}}, \bibinfo
  {author} {\bibfnamefont{V.}~\bibnamefont{Dexheimer}},\ and\ \bibinfo {author}
  {\bibfnamefont{S.}~\bibnamefont{Schramm}},\ }%
  \bibfield{journal}{%
  \bibinfo {journal} {Physical Review C}\ }%
  \textbf{\bibinfo {volume} {82}},\ \bibinfo {pages} {1} (\bibinfo {year}
  {2010})%
  \bibAnnoteFile{NoStop}{Negreiros2010}%
\bibitem{Dexheimer2010}%
  \BibitemOpen
  \bibfield{author}{%
  \bibinfo {author} {\bibfnamefont{V.~A.}\ \bibnamefont{Dexheimer}}\ and\
  \bibinfo {author} {\bibfnamefont{S.}~\bibnamefont{Schramm}},\ }%
  \bibfield{journal}{%
  \bibinfo {journal} {Physical Review C}\ }%
  \textbf{\bibinfo {volume} {81}} (\bibinfo {year} {2010})%
  \bibAnnoteFile{NoStop}{Dexheimer2010}%
\bibitem{Glendenning1989}%
  \BibitemOpen
  \bibfield{author}{%
  \bibinfo {author} {\bibfnamefont{N.~K.}\ \bibnamefont{Glendenning}},\ }%
  \bibfield{journal}{%
  \bibinfo {journal} {Nuclear Physics A}\ }%
  \textbf{\bibinfo {volume} {493}},\ \bibinfo {pages} {521} (\bibinfo {year}
  {1989})%
  \bibAnnoteFile{NoStop}{Glendenning1989}%
\bibitem{Chodos1974}%
  \BibitemOpen
  \bibfield{author}{%
  \bibinfo {author} {\bibfnamefont{A.}~\bibnamefont{Chodos}}, \bibinfo {author}
  {\bibfnamefont{R.}~\bibnamefont{Jaffe}}, \bibinfo {author}
  {\bibfnamefont{K.}~\bibnamefont{Johnson}}, \bibinfo {author}
  {\bibfnamefont{C.}~\bibnamefont{Thorn}},\ and\ \bibinfo {author}
  {\bibfnamefont{V.}~\bibnamefont{Weisskopf}},\ }%
  \bibfield{journal}{%
  \bibinfo {journal} {Physical Review D}\ }%
  \textbf{\bibinfo {volume} {9}},\ \bibinfo {pages} {3471} (\bibinfo {year}
  {1974})%
  \bibAnnoteFile{NoStop}{Chodos1974}%
\bibitem{Chodos1974a}%
  \BibitemOpen
  \bibfield{author}{%
  \bibinfo {author} {\bibfnamefont{A.}~\bibnamefont{Chodos}}, \bibinfo {author}
  {\bibfnamefont{R.}~\bibnamefont{Jaffe}}, \bibinfo {author}
  {\bibfnamefont{K.}~\bibnamefont{Johnson}},\ and\ \bibinfo {author}
  {\bibfnamefont{C.}~\bibnamefont{Thorn}},\ }%
  \bibfield{journal}{%
  \bibinfo {journal} {Physical Review D}\ }%
  \textbf{\bibinfo {volume} {10}},\ \bibinfo {pages} {2599} (\bibinfo {year}
  {1974})%
  \bibAnnoteFile{NoStop}{Chodos1974a}%
\bibitem{Farhi1984}%
  \BibitemOpen
  \bibfield{author}{%
  \bibinfo {author} {\bibfnamefont{E.}~\bibnamefont{Farhi}}\ and\ \bibinfo
  {author} {\bibfnamefont{R.}~\bibnamefont{Jaffe}},\ }%
  \bibfield{journal}{%
  \bibinfo {journal} {Physical Review D}\ }%
  \textbf{\bibinfo {volume} {30}},\ \bibinfo {pages} {2379} (\bibinfo {year}
  {1984})%
  \bibAnnoteFile{NoStop}{Farhi1984}%
\bibitem{Weber2005}%
  \BibitemOpen
  \bibfield{author}{%
  \bibinfo {author} {\bibfnamefont{F.}~\bibnamefont{Weber}},\ }%
  \bibfield{journal}{%
  \bibinfo {journal} {Progress in Particle and Nuclear Physics}\ }%
  \textbf{\bibinfo {volume} {54}},\ \bibinfo {pages} {193} (\bibinfo {year}
  {2005})%
  \bibAnnoteFile{NoStop}{Weber2005}%
\bibitem{Alcock1986}%
  \BibitemOpen
  \bibfield{author}{%
  \bibinfo {author} {\bibfnamefont{C.}~\bibnamefont{Alcock}}, \bibinfo {author}
  {\bibfnamefont{E.}~\bibnamefont{Farhi}},\ and\ \bibinfo {author}
  {\bibfnamefont{A.}~\bibnamefont{Olinto}},\ }%
  \bibfield{journal}{%
  \bibinfo {journal} {The Astrophysical Journal}\ }%
  \textbf{\bibinfo {volume} {310}},\ \bibinfo {pages} {261} (\bibinfo {year}
  {1986})%
  \bibAnnoteFile{NoStop}{Alcock1986}%
\bibitem{Usov2004}%
  \BibitemOpen
  \bibfield{author}{%
  \bibinfo {author} {\bibfnamefont{V.}~\bibnamefont{Usov}},\ }%
  \bibfield{journal}{%
  \bibinfo {journal} {Physical Review D}\ }%
  \textbf{\bibinfo {volume} {70}},\ \bibinfo {pages} {14} (\bibinfo {year}
  {2004})%
  \bibAnnoteFile{NoStop}{Usov2004}%
\bibitem{Negreiros2009}%
  \BibitemOpen
  \bibfield{author}{%
  \bibinfo {author} {\bibfnamefont{R.}~\bibnamefont{Negreiros}}, \bibinfo
  {author} {\bibfnamefont{F.}~\bibnamefont{Weber}}, \bibinfo {author}
  {\bibfnamefont{M.}~\bibnamefont{Malheiro}},\ and\ \bibinfo {author}
  {\bibfnamefont{V.}~\bibnamefont{Usov}},\ }%
  \bibfield{journal}{%
  \bibinfo {journal} {Physical Review D}\ }%
  \textbf{\bibinfo {volume} {80}},\ \bibinfo {pages} {083006} (\bibinfo {year}
  {2009})%
  \bibAnnoteFile{NoStop}{Negreiros2009}%
\bibitem{Negreiros2010b}
R. P. Negreiros,\ I. Mishustin,\ S. Schramm,\ \& F. Weber, Physical Review D {\bf 82},
103010 (2010).
\bibitem{Alford2001}%
  \BibitemOpen
  \bibfield{author}{%
  \bibinfo {author} {\bibfnamefont{M.}~\bibnamefont{Alford}},\ }%
  \bibfield{journal}{%
  \bibinfo {journal} {Annual Review of Nuclear and Particle Science}\ }%
  \textbf{\bibinfo {volume} {51}},\ \bibinfo {pages} {131} (\bibinfo {year}
  {2001})%
  \bibAnnoteFile{NoStop}{Alford2001}%
\bibitem{Alford2008}%
  \BibitemOpen
  \bibfield{author}{%
  \bibinfo {author} {\bibfnamefont{M.}~\bibnamefont{Alford}}, \bibinfo {author}
  {\bibfnamefont{A.}~\bibnamefont{Schmitt}}, \bibinfo {author}
  {\bibfnamefont{K.}~\bibnamefont{Rajagopal}},\ and\ \bibinfo {author}
  {\bibfnamefont{T.}~\bibnamefont{Sch\"{a}fer}},\ }%
  \bibfield{journal}{%
  \bibinfo {journal} {Reviews of Modern Physics}\ }%
  \textbf{\bibinfo {volume} {80}},\ \bibinfo {pages} {1455} (\bibinfo {year}
  {2008})%
  \bibAnnoteFile{NoStop}{Alford2008}%
\bibitem{Alford2001a}%
  \BibitemOpen
  \bibfield{author}{%
  \bibinfo {author} {\bibfnamefont{M.}~\bibnamefont{Alford}}, \bibinfo {author}
  {\bibfnamefont{J.}~\bibnamefont{Bowers}},\ and\ \bibinfo {author}
  {\bibfnamefont{K.}~\bibnamefont{Rajagopal}},\ }%
  \bibfield{journal}{%
  \bibinfo {journal} {Physical Review D}\ }%
  \textbf{\bibinfo {volume} {63}},\ \bibinfo {pages} {074016} (\bibinfo {year}
  {2001})%
  \bibAnnoteFile{NoStop}{Alford2001a}%
\bibitem{Bowers2002}%
  \BibitemOpen
  \bibfield{author}{%
  \bibinfo {author} {\bibfnamefont{J.}~\bibnamefont{Bowers}}\ and\ \bibinfo
  {author} {\bibfnamefont{K.}~\bibnamefont{Rajagopal}},\ }%
  \bibfield{journal}{%
  \bibinfo {journal} {Physical Review D}\ }%
  \textbf{\bibinfo {volume} {66}},\ \bibinfo {pages} {065002} (\bibinfo {year}
  {2002})%
  \bibAnnoteFile{NoStop}{Bowers2002}%
\bibitem{Lugones2002}%
  \BibitemOpen
  \bibfield{author}{%
  \bibinfo {author} {\bibfnamefont{G.}~\bibnamefont{Lugones}}\ and\ \bibinfo
  {author} {\bibfnamefont{J.~E.}\ \bibnamefont{Horvath}},\ }%
  \bibfield{journal}{%
  \bibinfo {journal} {Physical Review D}\ }%
  \textbf{\bibinfo {volume} {66}},\ \bibinfo {pages} {1} (\bibinfo {year}
  {2002})%
  \bibAnnoteFile{NoStop}{Lugones2002}%
\bibitem{Alford2003}%
  \BibitemOpen
  \bibfield{author}{%
  \bibinfo {author} {\bibfnamefont{M.}~\bibnamefont{Alford}}\ and\ \bibinfo
  {author} {\bibfnamefont{S.}~\bibnamefont{Reddy}},\ }%
  \bibfield{journal}{%
  \bibinfo {journal} {Physical Review D}\ }%
  \textbf{\bibinfo {volume} {67}},\ \bibinfo {pages} {1} (\bibinfo {year}
  {2003})%
  \bibAnnoteFile{NoStop}{Alford2003}%
\bibitem{Baym1971}%
  \BibitemOpen
  \bibfield{author}{%
  \bibinfo {author} {\bibfnamefont{G.}~\bibnamefont{Baym}}, \bibinfo {author}
  {\bibfnamefont{C.}~\bibnamefont{Pethick}},\ and\ \bibinfo {author}
  {\bibfnamefont{P.}~\bibnamefont{Sutherland}},\ }%
  \bibfield{journal}{%
  \bibinfo {journal} {The Astrophysical Journal}\ }%
  \textbf{\bibinfo {volume} {170}},\ \bibinfo {pages} {299} (\bibinfo {year}
  {1971})%
  \bibAnnoteFile{NoStop}{Baym1971}%
\bibitem{Gudmundsson1982}%
  \BibitemOpen
  \bibfield{author}{%
  \bibinfo {author} {\bibfnamefont{E.~H.}\ \bibnamefont{Gudmundsson}}, \bibinfo
  {author} {\bibfnamefont{C.~J.}\ \bibnamefont{Pethick}},\ and\ \bibinfo
  {author} {\bibfnamefont{R.~I.}\ \bibnamefont{Epstein}},\ }%
  \bibfield{journal}{%
  \bibinfo {journal} {The Astrophysical Journal}\ }%
  \textbf{\bibinfo {volume} {259}},\ \bibinfo {pages} {L19} (\bibinfo {year}
  {1982})%
  \bibAnnoteFile{NoStop}{Gudmundsson1982}%
\bibitem{Gudmundsson1983}%
  \BibitemOpen
  \bibfield{author}{%
  \bibinfo {author} {\bibfnamefont{R.~I.~E.}\ \bibnamefont{{E. H. Gudmundsson,
  C. J. Pethick}}},\ }%
  \bibfield{journal}{%
  \bibinfo {journal} {The Astrophysical Journal}\ }%
  \textbf{\bibinfo {volume} {272}},\ \bibinfo {pages} {286} (\bibinfo {year}
  {1983})%
  \bibAnnoteFile{NoStop}{Gudmundsson1983}%
\bibitem{Yakovlev2001a}%
  \BibitemOpen
  \bibfield{author}{%
  \bibinfo {author} {\bibfnamefont{D.}~\bibnamefont{Yakovlev}}, \bibinfo
  {author} {\bibfnamefont{A.~D.}\ \bibnamefont{Kaminker}}, \bibinfo {author}
  {\bibfnamefont{O.~Y.}\ \bibnamefont{Gnedin}},\ and\ \bibinfo {author}
  {\bibfnamefont{P.}~\bibnamefont{Haensel}},\ }%
  \bibfield{journal}{%
  \bibinfo {journal} {Physics Reports}\ }%
  \textbf{\bibinfo {volume} {354}},\ \bibinfo {pages} {1} (\bibinfo {year}
  {2001})%
  \bibAnnoteFile{NoStop}{Yakovlev2001a}%
\bibitem{IWAMOTO1982}%
  \BibitemOpen
  \bibfield{author}{%
  \bibinfo {author} {\bibfnamefont{N.}~\bibnamefont{Iwamoto}},\ }%
  \bibfield{journal}{%
  \bibinfo {journal} {Annals of Physics}\ }%
  \textbf{\bibinfo {volume} {141}},\ \bibinfo {pages} {1} (\bibinfo {year}
  {1982})%
  \bibAnnoteFile{NoStop}{IWAMOTO1982}%
\bibitem{Flowers1981}%
  \BibitemOpen
  \bibfield{author}{%
  \bibinfo {author} {\bibfnamefont{E.}~\bibnamefont{Flowers}}\ and\ \bibinfo
  {author} {\bibfnamefont{N.}~\bibnamefont{Itoh}},\ }%
  \bibfield{journal}{%
  \bibinfo {journal} {The Astrophysical Journal}\ }%
  \textbf{\bibinfo {volume} {250}},\ \bibinfo {pages} {750} (\bibinfo {year}
  {1981})%
  \bibAnnoteFile{NoStop}{Flowers1981}%
\bibitem{Haensel1991}%
  \BibitemOpen
  \bibfield{author}{%
  \bibinfo {author} {\bibfnamefont{P.}~\bibnamefont{Haensel}},\ }%
  \bibfield{journal}{%
  \bibinfo {journal} {Nuclear Physics B - Proceedings Supplements}\ }%
  \textbf{\bibinfo {volume} {24}},\ \bibinfo {pages} {23} (\bibinfo {year}
  {1991})%
  \bibAnnoteFile{NoStop}{Haensel1991}%
\bibitem{Espinoza2011}
C. M. Espinoza,\ a. G. Lyne,\ M. Kramer,\ R. N. Manchester,\ \& V. M. Kaspi, The Astrophysical
Journal {\bf 741}, L13 (2011).  
  \bibitem{Kurkela2010}%
  \BibitemOpen
  \bibfield{author}{%
  \bibinfo {author} {\bibfnamefont{A.}~\bibnamefont{Kurkela}}, \bibinfo
  {author} {\bibfnamefont{P.}~\bibnamefont{Romatschke}}, \bibinfo {author}
  {\bibfnamefont{A.}~\bibnamefont{Vuorinen}},\ and\ \bibinfo {author}
  {\bibfnamefont{B.}~\bibnamefont{Wu}},\ }%
  \bibfield{journal}{%
  \bibinfo {journal} {arXiv/1006.4062}}%
   (\bibinfo {year} {2010})%
  \bibAnnoteFile{NoStop}{Kurkela2010}%
\bibitem{Blaschke2010}%
  \BibitemOpen
  \bibfield{author}{%
  \bibinfo {author} {\bibfnamefont{D.}~\bibnamefont{Blaschke}}, \bibinfo
  {author} {\bibfnamefont{J.}~\bibnamefont{Berdermann}},\ and\ \bibinfo
  {author} {\bibfnamefont{R.}~\bibnamefont{Lastowiecki}},\ }%
  \bibfield{journal}{%
  \bibinfo {journal} {arXiv/1009.1181}}%
   (\bibinfo {year} {2010})%
  \bibAnnoteFile{NoStop}{Blaschke2010}%
  \bibitem{Demorest2011}
P. B. Demorest,\ T. Pennucci,\ S. M. Ransom,\ M. S. E. Roberts,\ \& J. W. T.
Hessels,\ Nature
{\bf 467}, 1081 (2010).
\bibitem{Page2011a}
D. Page,\ M. Prakash,\ J. Lattimer,\ \& A. Steiner, Physical Review Letters {\bf 106}, 081101
(2011).  
\bibitem{Yakovlev2011a}
P. S. Shternin,\ D. G. Yakovlev,\ C. O. Heinke,\ W. C. G. Ho,\ \& D. J. Patnaude,
Monthly Notices of the Royal Astronomical Society: Letters {\bf 412}, L108-L112 (2011).
\bibitem{Noda2011}
T. Noda, M. Hashimoto, Y. Matsuo et al, astro-ph/1109.1080
\bibitem{Negreiros2011}
R. Negreiros, S. Schreamm, \& F. Weber, astro-ph/1103.3870

\end{thebibliography}
\end{document}